\documentclass{iopart}

\usepackage[english]{babel}
\usepackage{graphicx}
\usepackage{amssymb}
\usepackage{bm}
\usepackage{psfrag}

\begin{document}

\title{Quantum dislocations: the fate of multiple vacancies in two dimensional solid $^4$He}

\author{M. Rossi, E. Vitali\footnote{Present address:
INFM--CNR DEMOCRITOS National Simulation Center,via Beirut 2-4, 34014 Trieste, Italy},
D.E. Galli and L. Reatto}
 \address{Dipartimento di Fisica, Universit\`a degli Studi
          di Milano, via Celoria 16, 20133 Milano, Italy}
 \ead{maurizio.rossi@unimi.it}

\date{\today}

\begin{abstract}
Defects are believed to play a fundamental role in the supersolid state of $^4$He.
We have studied solid $^4$He in two dimensions (2D) as function of the number of vacancies $n_v$,
up to 30, inserted in the initial configuration at $\rho = 0.0765$\AA$^{-2}$, close to the melting
density, with the exact zero temperature Shadow Path Integral Ground State method.
The crystalline order is found to be stable also in presence of many vacancies and we observe two
completely different regimes.
For small $n_v$, up to about 6, vacancies form a bound state and cause a decrease of the 
crystalline order.
At larger $n_v$, the formation energy of an extra vacancy at fixed density decreases by one order
of magnitude to about 0.6 K.
In the equilibrated state it is no more possible to recognize vacancies because they mainly 
transform into quantum dislocations and crystalline order is found almost independent on how many 
vacancies have been inserted in the initial configuration.
The one--body density matrix in this latter regime shows a non decaying large distance tail:
dislocations, that in 2D are point defects, turn out to be mobile, their number is fluctuating,
and they are able to induce exchanges of particles across the system mainly triggered by the 
dislocation cores.
These results indicate that the notion of incommensurate versus commensurate state loses meaning
for solid $^4$He in 2D, because the number of lattice sites becomes ill defined when the system
is not commensurate. 
Crystalline order is found to be stable also in 3D in presence of up to 100 vacancies.
\end{abstract}

 %Solid helium and related quantum crystal

\maketitle

\section{INTRODUCTION}
\label{sec:intr}

Solid $^4$He is the subject of many experimental and theoretical studies\cite{Prok,Bali,Rev} 
since the discovery of non classical rotational inertia\cite{Chan} (NCRI), an expected 
manifestation of supersolidity.\cite{Legg} 
Supersolidity is a new state of matter in which spatial order and off--diagonal long range order
(ODLRO) are present at the same time, implying some form of superfluid properties, and was 
already predicted long ago.\cite{Andr,Ches} 
This novel state of matter attracts interest also in the prospect of finding it in cold atoms in
optical lattices.\cite{Scar} 
However the precise nature of solid $^4$He at low temperature is still elusive.
Experimental evidence suggests that defects play an important role in NCRI, but which kind of 
disorder is responsible for the anomalous properties of solid $^4$He is still not clear.
Many different defects have been considered, but none of the proposed models has shown to be able
to capture all the phenomenology of supersolidity.
For example the stiffening of $^4$He below the ``transition'' temperature\cite{Beam} suggests
dislocations as a candidate, but this possibility has difficulty in explaining the NCRI seen in
solid $^4$He in Vycor\cite{Chan} or in Aerogel.\cite{Chan2}
Grain boundaries have been considered too,\cite{Prok} but NCRI has been observed also in single 
crystals.\cite{Chan3}

Defected solid $^4$He systems have been studied by means of Quantum Monte Carlo (QMC)
techniques.\cite{Pede,Gallo,Gall1,spigs,Gall,Boni,Cepe2,Pollet,Pollet2,Boni2,Corb,Soyl}
Such QMC methods are by construction equilibrium methods, so defects have to be stabilized by 
periodic boundary conditions (via a suitable choice of the simulation box) or by fixing the 
degrees of freedom of a number of atoms surrounding the interested defect.
This constraint on the configurational space usually does not prevent the study of the physical 
properties of the defected system, and the results are generally in a good quantitative agreement
with experimental observations; see, for example, the estimation of the vacancy activation 
energy\cite{Pede} or the binding energy to the dislocation core of a $^3$He atom.\cite{Corb} 
Exact QMC results, both at finite\cite{Boni,Cepe2,Pollet} and zero temperature\cite{Ross2},
show that 2 and 3 vacancies form a bound state.
A systematic study of many vacancies in solid $^4$He is still lacking; questions like whether the
crystalline structure is stable in presence of a large number of vacancies, whether the 
vacancy--vacancy interaction is pairwise additive (or almost so), whether many vacancies lead to 
phase separation as suggested in Ref.~\cite{Boni,Pollet}, or whether vacancies turn %
themselves into other kinds of defects are, according to us, still unanswered.

Since the earliest days of supersolidity quantum lattice models have been considered.\cite{Liu} 
A terminology borrowed from lattice models is currently used for solid $^4$He in terms of a
commensurate state (i.e. a perfect crystal in which there is an integer number of atoms per unit 
cell) or an incommensurate one (a crystal with non integer probability of occupation of the 
unit cell).
QMC at $T=0$ K of solid $^4$He with one or few vacancies in the simulation cell beautifully\cite{Gall}
confirms the picture of an incommensurate state: the vacancy(-ies) is (are) mobile and the 
number of density maxima is equal to the number of particles {\it plus} the number of vacancies.
But are we sure that we understand vacancies at a finite concentration in a macroscopic system?
Is the option commensurate or incommensurate really appropriate for solid $^4$He?
One should keep in mind that in a lattice model or in cold atoms in an optical lattice the 
lattice periodicity is externally imposed, whereas in solid $^4$He the same dynamical entities, 
the atoms, have to build up the periodicity as well as the delocalization required for 
supersolidity.
Such ``mermaid'' aspect of the atoms is unique to a quantum solid.
The purpose of this article is to address such issues with exact QMC methods at $T=0$ K in two 
dimensions (2D).

We find that size and box commensuration effects are very pronounced, so in order to be able to 
explore very large distances in a Monte Carlo calculation, up to 100 \AA, we have mainly studied
solid $^4$He in 2D.    
In this context, the 2D solid is also of interest as a simple model for out of registry solid 
$^4$He adsorbed on planar substrates, such as graphite or silica glasses.
New experimental investigations\cite{Trieste} are under way to answer the question whether a 
supersolid state is present also in adsorbed $^4$He.
We have studied 2D solid $^4$He systems at $T=0$ K with a number of atoms up and above thousand 
containing up to 30 vacancies.
We find that crystalline order is stable also in presence of a large number of vacancies, at 
least in the range $\lesssim 2.5$\% of vacancy concentration studied by us.
Multiple vacancies are highly correlated with a preference to form linear structures, but in 
presence of 10 or more vacancies the scars in the lattice are healed away transforming the 
vacancies mainly into dislocations.
Such dislocations are not permanent structures, but are mobile and their number is fluctuating.
This allows exchange of particles across the system, not only in the cores of dislocations, 
giving the possibility of establishing a well defined, at least local, phase\cite{Ander}
and indeed, in the simulated systems, we find a non decaying large distance tail in the
one--body density matrix.
This hints at the possibility that an extended system is supersolid but more work is needed 
to corroborate this conclusion.

We do not address here the origin of vacancies/dislocations, i.e. if the true ground state of 
solid $^4$He contains or not zero--point defects.
Our finding is that solid $^4$He, at least in 2D, is much more complex than the picture derived 
from the notion of commensurate vs incommensurate state because the number of lattice sites is 
an ill defined quantity when dislocations are present.
The number of lattice sites is a meaningful quantity, at least for the sizes we are able to 
simulate, only if the number of particles and the simulation box are such that the regular 
lattice exactly fits in or the misfit is limited to just one or very few vacancies.
We have performed few simulations also in three dimensions (3D).
Also in this case we find that, near melting, crystalline order is stable even in presence of 
100 vacancies, but we have not yet performed a detailed characterization of the disorder
in the system.

The paper is organized as follows: Sec.~\ref{sec:meth} deals with the exact $T=0$ K SPIGS method.
Details on the simulations are outlined in Sec.~\ref{sec:deta}.
Sec.~\ref{sec:resu} contains our results and our conclusions are given in Sec.~\ref{sec:conc}.

\section{THE SPIGS METHOD}
\label{sec:meth}

We employ the exact $T=0$ K Shadow Path Integral Ground State (SPIGS)\cite{spigs} method, an 
``exact'' QMC technique.\cite{Vita,Ross3}
SPIGS is an extension of the Path Integral Ground State (PIGS)\cite{pigs} method.
The aim of PIGS is to improve a variationally optimized trial wave function $\psi_T$ by 
constructing a path in the Hilbert space of the system which connects the given $\psi_T$ to the 
lowest energy state of the system, $\psi_0$, constrained by the choice of the number of 
particle $N$, the geometry and the
boundary conditions of the simulation box and the density $\rho$.
During this ``path'', the correct correlations among the particles arise through the ``imaginary
time evolution operator'' $e^{-\tau\hat H}$, where $\hat H$ is the Hamiltonian operator.
For a large enough $\tau$, an accurate representation for the lowest energy state wave function 
is given by $\psi_\tau=e^{-\tau\hat H}\psi_T$, which can be written analytically by discretizing 
the path in imaginary time and exploiting the factorization property
$e^{-(\tau_1+\tau_2)\hat H} =e^{-\tau_1\hat H}e^{-\tau_2\hat H}$.
In this way, $\psi_\tau$ turns out to be expressed in terms of convolution integrals which involve
the ``imaginary time propagator'' $\langle R|e^{-\delta\tau\hat H}|R'\rangle$ for small 
$\delta\tau$, for which very accurate approximations can be found in literature.\cite{Cepe,Suzu}
This maps the quantum system into a classical system of open polymers.\cite{pigs}
An appealing feature peculiar to the PIGS method is that, in $\psi_\tau$, the variational ansatz
acts only as a starting point, while the full path in imaginary time is governed by 
$e^{-\tau\hat H}$, which depends only on the Hamiltonian operator.
We have recently shown that PIGS results for large enough $\tau$ are unaffected by the choice of
$\psi_T$ both in the liquid and in the solid phase\cite{Vita,Ross3} thus providing an unbiased 
exact $T=0$ K QMC method.
Within SPIGS a shadow wave function (SWF)\cite{Viti,Moro} is taken as $\psi_T$ and this choice
was shown to greatly accelerate convergence to $\psi_0$.
Another appealing feature of the SPIGS method is that it recovers the solid phase via a 
spontaneously broken translational symmetry\cite{spigs} and there is no constriction on the 
atomic positions, so it is particularly indicated in studying crystals with defects. 

\section{MODEL SYSTEM AND SIMULATION DETAILS}
\label{sec:deta}

By studying the equation of state we find the freezing and the melting densities to be, 
respectively, $\rho_f=0.0672$ \AA$^{-2}$ and $\rho_m=0.0724$ \AA$^{-2}$.
As simulation box we choose then a rectangular box fitting a regular triangular lattice with $M$ 
lattice sites at $\rho = 0.0765$ \AA$^{-2}$.
Periodic boundary conditions (pbc) are applied in both directions.
Dealing with low temperature properties, $^4$He atoms are described as structureless zero--spin 
bosons, interacting through the HFDHE2 Aziz potential.\cite{Aziz}
In order to avoid any approximation associated with the estimation of tail corrections due to the
finite size of the simulation boxes, we have considered a truncated and shifted Aziz potential 
which goes to zero at $r_{\rm cut}=6$ \AA.
Since $r_{\rm cut}$ is well below the minimum size of each considered box, the results relative 
to different box sizes can be compared without any further correction.
Our SPIGS computation is based on the pair--product approximation for the imaginary time 
projector\cite{Cepe} and the imaginary time step $\delta\tau=1/40$ K$^{-1}$ has been chosen in 
order to ensure a good accuracy and a reasonable computational effort.\cite{Vita}

\begin{figure}
 \begin{center}
  \includegraphics[width=8.5cm]{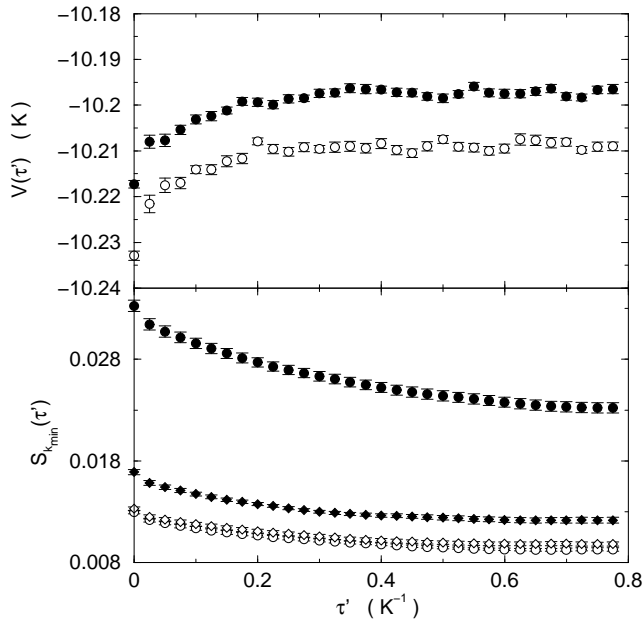}
 \end{center}
 \caption{\label{f:lr} Upper panel: Mixed expectation value for the potential energy per particle as
          a function of the projection time $\tau'$ computed in a 2D $^4$He crystals at
          $\rho = 0.0765$\AA$^{-2}$  in the $M=960$ box, with $n_v=0$ (open symbols) and $n_v=20$
          (filled symbols).
          Lower panel: Static structure factor computed in the same systems for the lowest
          wave-vectors along the first neighbor direction (circles, $k_{\rm min}=0.055$\AA$^{-1}$) 
          and orthogonal to the first neighbor direction (diamonds, $k_{\rm min}=0.059$\AA$^{-1}$).
          Open and filled symbols have the same meaning ad in upper panel.}
\end{figure}

An important parameter in the computation is the total projection time $\tau$.
Any trial wave function $\psi_T$ can be expanded in terms of the exact ground state and of the
excited states of zero total linear momentum.
If we call $\Delta\varepsilon_{\rm min}$ the minimum excitation energy for such excited states, 
$\tau$ has to be larger than $(\Delta\varepsilon_{\rm min})^{-1}$ in order to project out any 
excitation contribution in $\psi_T$; the rate of rejection of such contributions depends also on 
the quantity one is computing due to sensitivity to missing long range correlations in $\psi_T$ 
or to wrong short range correlations.
In the perfect crystal, using K as energy unit, $\tau=0.775$ K$^{-1}$ was found\cite{Vita} to
be large enough to ensure convergence of both diagonal and off--diagonal observables when a SWF 
is used as $\psi_T$.
This is compatible with the criterion $\tau>(\Delta\varepsilon_{\rm min})^{-1}$.
In fact, in a perfect crystal, the lowest energy excitations are phonons; thus 
$\Delta\varepsilon_{\rm min}$ corresponds to the combination of two phonons of opposite $\vec k$
with the smallest achievable $k$ value, i.e. in a box with pbc $k_{\rm min}=2\pi/L_{\rm max}$ 
where $L_{\rm max}$ is the largest side.
Via a new analytic continuation method\cite{gift} we have estimated the longitudinal sound 
velocity for 2D solid $^4$He at $\rho = 0.0765$\AA$^{-2}$ to be $c=36.5 \pm 2.5$ K\AA.
Then the minimum imaginary time requested for convergence is 
$\tau_{\rm min}=1/\Delta\varepsilon_{\rm min}\simeq 0.25$ K$^{-1}$ in the largest simulation 
box considered here.
This value is well below the used one $\tau=0.775$ K$^{-1}$.
Since the transverse sound velocity is typically similar to $c$, the considered $\tau$ value is 
enough to ensure also the removal of transverse phonon contribution.
The situation might be different in presence of vacancies, or other kinds of defects, because
novel low energy excitation modes could be present. 
We have performed several tests in order to be confident that also for defected crystals the 
considered value of $\tau$ is large enough.
The most straightforward one is to perform simulations with larger $\tau$ values and verify that 
the expectation values are compatible within the statistical uncertainty.
Such a test turns out to be very time consuming and in practice it has been feasible only for
a number of $^4$He atom $N \lesssim 250$,  where we have increased the value of $\tau$ up to 
$1.16$ K$^{-1}$ finding good convergence for diagonal properties. 
An indirect test on the convergence is provided by the behavior of suitably chosen observables
during the imaginary time path.
If we define
\begin{equation}
 \label{mixed}
  \mathcal{O}(\tau') = \frac{\langle\psi_{\tau'}|\hat{\mathcal{O}}|\psi_{2\tau-\tau'}\rangle}
                            {\langle\psi_{\tau'}|\psi_{2\tau-\tau'}\rangle}
\end{equation}
for $0\le\tau'\le\tau$, when $\tau$ is large enough we have $\psi_{2\tau-\tau'}\simeq\psi_0$ 
and Eq.~\ref{mixed} gives the mixed expectation value
$\mathcal{O}(\tau')=\langle\psi_{\tau'}|\hat{\mathcal{O}}|\psi_0\rangle/\langle\psi_{\tau'}|\psi_0\rangle$
as a function of the partial imaginary time $\tau'$.
If $\mathcal{O}(\tau')$ has a plateau when $\tau' \to \tau$, we are guaranteed that $\tau$ is 
large enough for convergence and the plateau represents the ``exact'' expectation value of 
$\hat{\mathcal{O}}$.
A couple of examples are shown in Fig.\ref{f:lr} for the potential energy ($\hat{\mathcal{O}}=\hat V$)
and for the static structure factor at the lowest wave vectors 
($\hat{\mathcal{O}}=\hat\rho_{-k_{\rm min}} \hat\rho_{k_{\rm min}}$, where 
$\hat\rho_{\vec k}=\frac{1}{\sqrt{N}}\sum_{j=1}^N\exp{(i\vec k\cdot \vec r_j)}$ is the density
fluctuation operator) along the simulation box axis.
As far as the potential energy $V$ is concerned, excitation modes are rapidly projected out and
$\tau\simeq0.25$ K$^{-1}$ is enough to achieve convergence both in the perfect and in the defected
systems, as shown in Fig.\ref{f:lr} by the presence pf a plateau of $V(\tau')$ in both cases.
Larger values of the imaginary time are needed to reach convergence for quantities that depend on 
long--range correlations, such as the static structure factor at small wave vectors 
($S_{k_{\rm min}}$) .\cite{Reatto} 
In the perfect crystal, convergence for $S_{k_{\rm min}}$ is achieved at about $0.6$ K$^{-1}$, as 
shown in Fig.\ref{f:lr}.
In the defected crystal (especially when many vacancies are present as in the case considered in the
figure) the chosen $\tau=0.775$ K$^{-1}$ value is just enough to get convergence.
An interesting physical effect emerges from Fig.\ref{f:lr}, the static structure factor is strongly
affected by the presence of defects especially along the nearest neighbor direction.

Two independent simulations are needed in order to calculate the formation energy of $n_v$ 
vacancies, one with $N=M$ particles (the regular ideal crystal, dubbed also perfect crystal),
and one with $n_v$ less atoms (defected crystal). 
In this second case we say that we have $n_v=M-N$ vacancies, even if in general we can talk of 
vacancies only in the starting configuration of the simulation, where $n_v$ atoms are removed 
from the perfect lattice.
In fact, in SPIGS, like in PIMC, there is no constraint on the atomic positions, so that
vacancies are free to transform into different kinds of defects.
In presence of vacancies, after removing $n_v$ particles from the ideal starting configuration, 
we rescale the dimensions of the simulation box to reset the system to the original density.
This is performed to circumvent the need of correcting the energy due to density changes caused 
by the inclusion of vacancies.
The formation energy of $n_v$ vacancies is proportional to the difference\cite{Pede,nota} between 
the energy per particle of the defected crystal, $e(M-n_v)$, and of the perfect one, $e(M)$, i.e.
\begin{equation}
 \Delta E_{n_v}=(M-n_v)[e(M-n_v)-e(M)].
\end{equation}
When vacancies turn into dislocations this $\Delta E_{n_v}$ has the meaning of defect formation 
energy at fixed density.
One can consider also the formation energy $\Delta \widetilde{E}_{n_v}$ at a fixed lattice 
parameter; this is given by
\begin{equation}
 \Delta \widetilde{E}_{n_v}=\Delta E_{n_v}-n_v\mu,
\end{equation}
$\mu$ being the chemical potential, that for our system turns out to be $\mu=14.22\pm0.02$K.

We check the presence of crystalline order by monitoring the static structure factor for the 
presence of Bragg peaks.
In addition, we compute the particle coordination number, via a Delaunay triangulation\cite{Dela}
(DT) of the sampled configurations, in order to estimate the amount of disorder in the 
system.\cite{Ches2}
In a perfectly ordered 2D triangular crystal each atom is linked to 6 other atoms in the DT.
Atoms with number of coordination not equal to 6 are then a measure of the disorder in the
crystal.\cite{Ches2}
A conservation law exists for the coordination numbers:\cite{Ches2}
\begin{equation}
 \sum_{i=3}^5(6-i)N_i=\sum_{i+7}^\infty (i-6)N_i
\end{equation}
where $N_i$ is the number of $i-$coordinated atoms, so that we can consider only $N_i$ with 
$i<6$.
Therefore we take $\tilde X_d=2N_4+N_5$ as an estimate of disorder in the system (we never found
3-sided polygons, i.e. $N_3=0$).
In a perfect ($N=M$) 2D quantum crystal the coordination is 6 only on the average, in fact, due to 
the large zero point motion, fluctuations are present such that atoms do not have always 
coordination 6.
Thus, a more useful quantity measuring the net amount of disorder in a crystal with defects is
the difference between the observed  $\tilde X_d$ and that of the corresponding perfect crystal:
\begin{equation}
 X_d=(2N_4+N_5)^{n_v}_M-(2N_4+N_5)^{n_v=0}_M.
\end{equation}

When the number of vacancies $n_v$ is small, up to 6, we have also computed the vacancy--vacancy
correlation function $g_{vv}(r)$ by recording the relative distances among vacancies during the 
Monte Carlo sampling.
The determination of the vector positions of the vacancies in a crystalline configuration is
far from being trivial due to the large zero point motion, to high vacancy mobility and because
in our algorithm the center of mass is not fixed.
Vacancy positions are obtained via the coarse--graining procedure illustrated in 
Ref.~\cite{Ross2}. %
This method is efficient as long as we have few vacancies.
For large $n_v$ the efficiency in recognizing the position of vacancies becomes extremely poor 
(in fact, as we shall see, vacancies turn into dislocations), then is no more possible to 
compute $g_{vv}(r)$.
Different definitions of vacancy positions, like the one in Ref.~\cite{Cepe2}, give very %
similar results.\cite{nota2}

A special care has been devoted in ruling out possible metastability effects.
We have considered different starting configurations for vacancies, i.e. we have removed a number
$n_v$ of atoms forming a compact cluster or a linear cluster, or the removed atoms come from 
random lattice positions.
After long enough equilibration (our simulations are never shorter than $6\times 10^5$ Monte 
Carlo steps) we find agreeing results for systems with the same $n_v$ value.
In order to rule out possible finite size effects and pbc bias we have considered systems of 
increasing sizes ($M$ values).
Again we find compatible results for systems of different sizes and with equal $n_v$ value, 
so no appreciable finite size effects have been detected.
The only exception is the one--body density matrix that will be discussed below.
Furthermore we have considered also crystals with no principal crystalline axis parallel to the 
box sides obtaining once more the same results.
Moreover, we have considered systems where the starting configuration was obtained by removing 
$n_v$ particles from an equilibrated configuration of the perfect crystal and at the same time
the positions of one (two) line(-s) of atoms, parallel to one (both) box side(-s) were kept 
fixed during the Monte Carlo sampling.
Even in this cases, the results do not change, for instance, for large $n_v$ values, vacancies 
are found to turn into dislocations with the same features as in the fully mobile systems.

\section{RESULTS}
\label{sec:resu}

\begin{figure}
 \begin{center}
  \includegraphics[width=8.5cm]{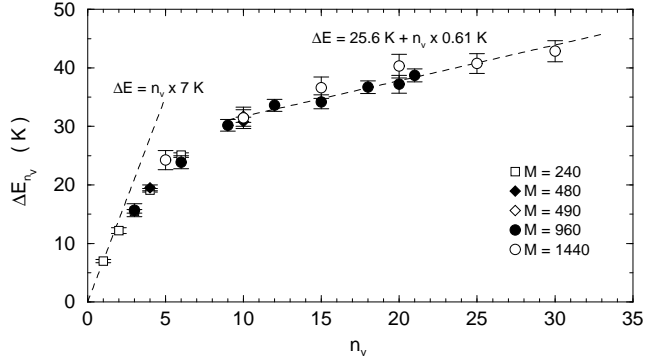}
 \end{center}
 \caption{\label{f:att} Defect formation energy $\Delta E_{n_v}$ as a function of $n_v$ 
          in 2D solid $^4$He at $\rho = 0.0765$\AA$^{-2}$ computed in boxes with different 
          lattice site numbers $M$. 
          Dashed lines are linear fit to the data.}
\end{figure}
\begin{figure}
 \begin{center}
  \includegraphics[width=8.5cm]{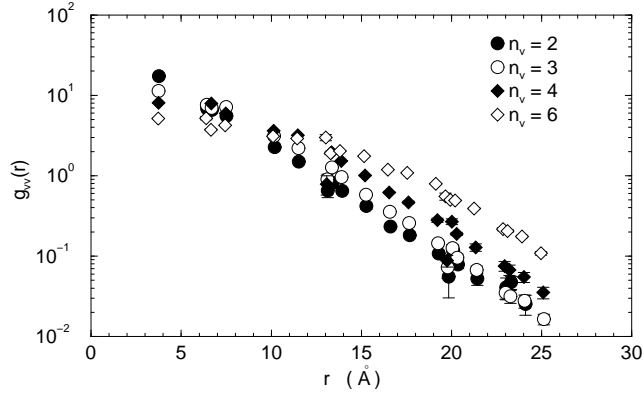}
 \end{center}
 \caption{\label{gvv_2D} Vacancy--vacancy correlation function $g_{vv}(r)$ computed in 2D solid
          $^4$He at $\rho=0.0765$\AA$^{-2}$ in a box with $M=240$ lattice sites for different
          number of vacancies $n_v$.
          $g_{vv}(r)$ is normalized with the correlation function of $n_v$ vacancies randomly
          distributed on the same lattice.}
\end{figure}
\begin{figure}
 \begin{center}
  \includegraphics[width=8.5cm]{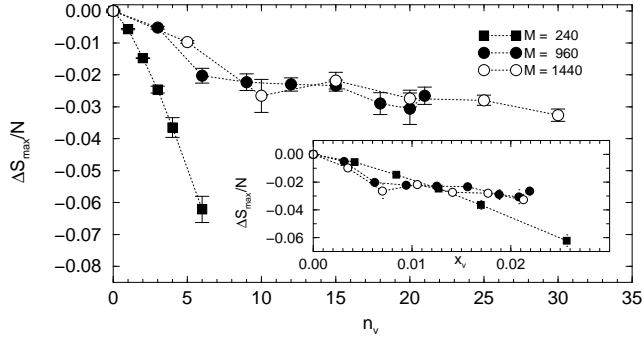}
 \end{center}
 \caption{\label{f:str} $\Delta S_{\rm max}/N=S^{n_v}_{\rm max}/(M-n_v)-S^{n_v=0}_{\rm max}/M$
          as a function of the number of vacancies $n_v$ (and of the concentration of vacancies
          $x_v$ in the inset) for different $M$.
          $S^{n_v}_{\rm max}$ is the main Bragg peak integrated intensity.\cite{nota3}
          The main Bragg peak is at $|\vec k|\simeq 1.87$\AA$^{-1}$.
          Lines are aid to the eyes.}
\end{figure}

We have studied a triangular crystal at $\rho = 0.0765$\AA$^{-2}$  with $M=240$, 480, 960 and 
1440 lattice positions of the perfect crystal.
Our results for the formation energy $\Delta E_{n_v}$ at constant density are plotted in 
Fig.~\ref{f:att}.
The dependence of $\Delta E_{n_v}$ on $n_v$ is monotonic and systematically sublinear, confirming
the existence of an attractive interaction among vacancies.
Systems with the same $n_v$ and different $M$ have the same $\Delta E_{n_v}$ within the 
statistical uncertainty, i.e. $\Delta E_{n_v}$ has no significant dependence on $x_v=n_v/M$, at 
least in the range $x_v \lesssim 2.5$ \% that we have studied.
From the plot in Fig.~\ref{f:att} it is possible to recognize two different behaviors: for 
$n_v<6$, $\Delta E_{n_v}$ deviates rapidly from the linear dependence and vacancies form a bound 
state as indicated by a (roughly) exponentially decreasing correlation function $g_{vv}(r)$,
plotted in Fig.~\ref{gvv_2D}.
For $n_v>10$ the ratio $\Delta E_{n_v} /n_v$ remains practically constant with a value of about 
0.61 K.
This means that, when some vacancies are already present in the system, the creation of an 
additional vacancy has a very low cost, about one tenth of the cost of a single vacancy.
Notice that in the linear regime  $\Delta E_{n_v}$, at fixed lattice parameter, is negative, there 
is a {\it gain} of 13.61 K for each additional vacancy.

\begin{figure}
 \begin{center}$
  \begin{array}{cc}
   \includegraphics[width=8.0cm]{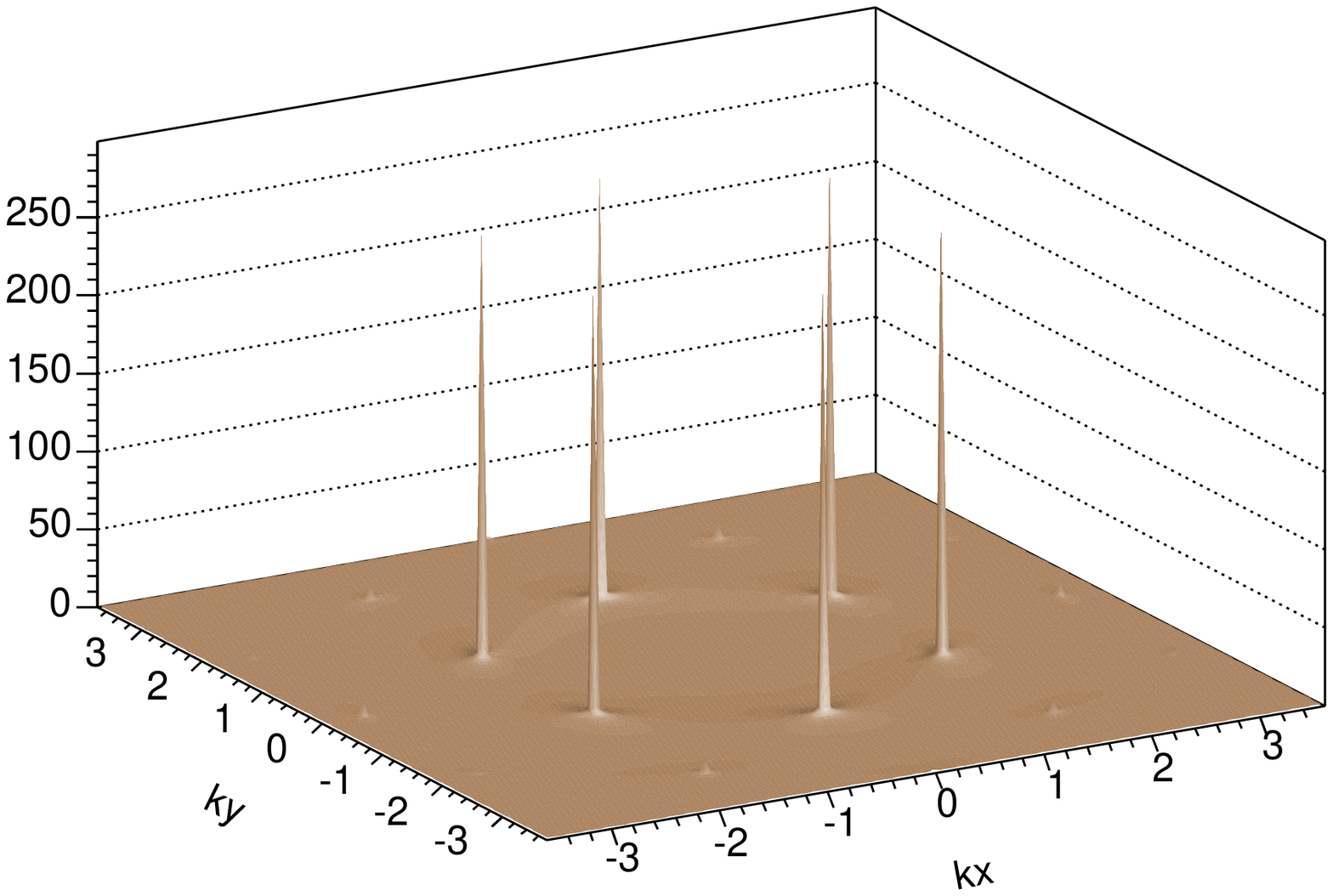} & \includegraphics[width=8.0cm]{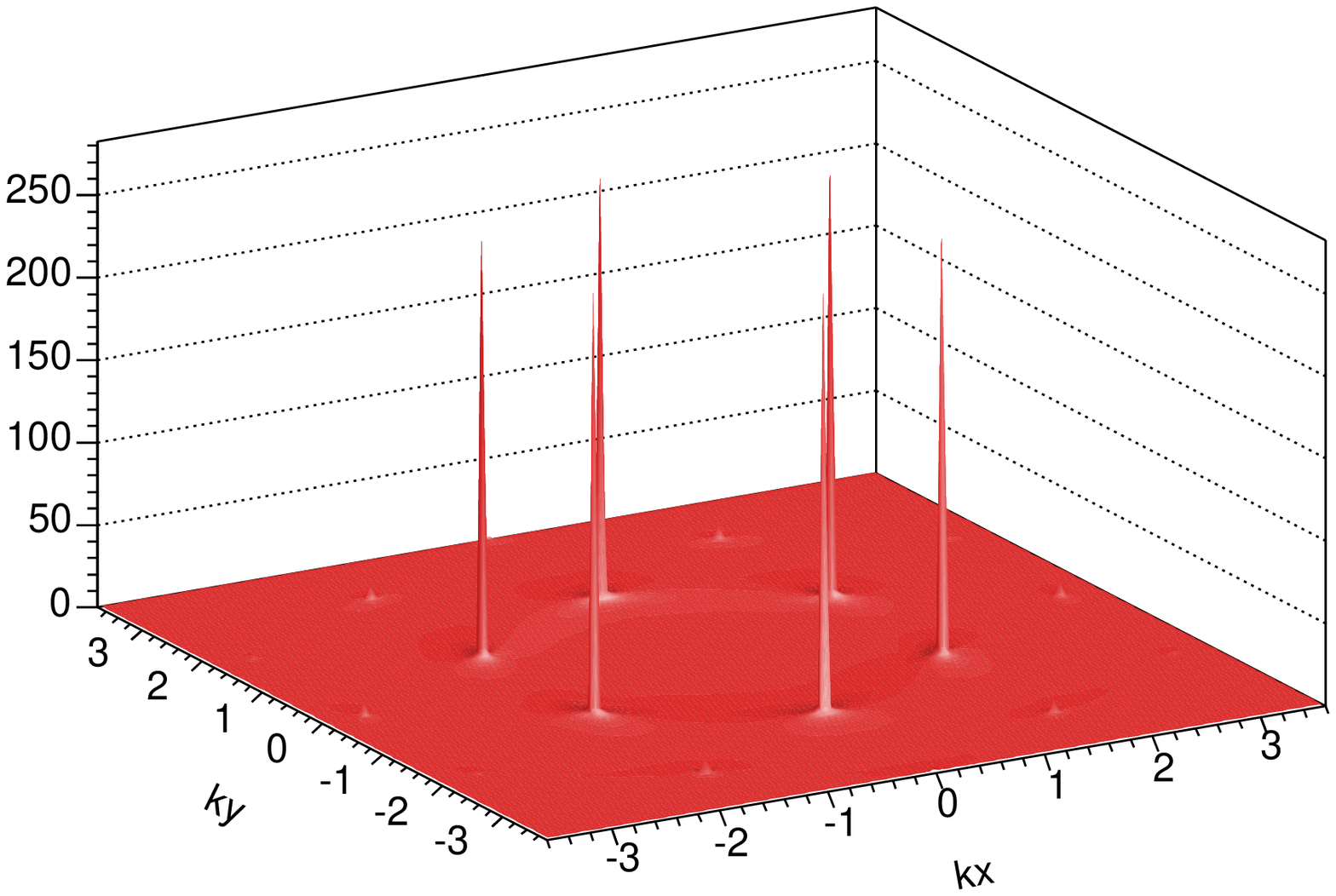} \\
   {\rm (a)}\qquad n_v= 0                   & {\rm (b)}\qquad n_v= 5                   \\
   \includegraphics[width=8.0cm]{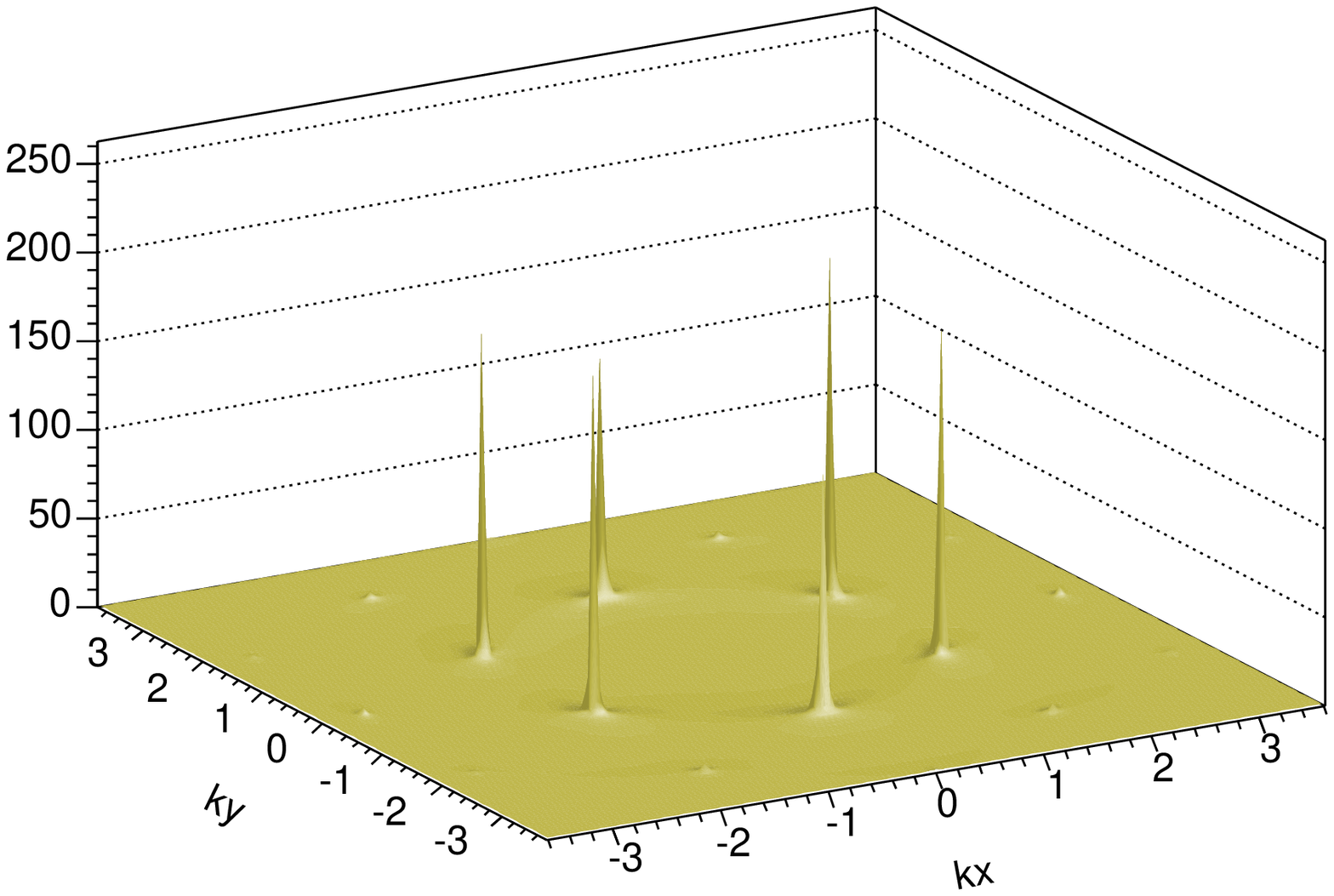} & \includegraphics[width=8.0cm]{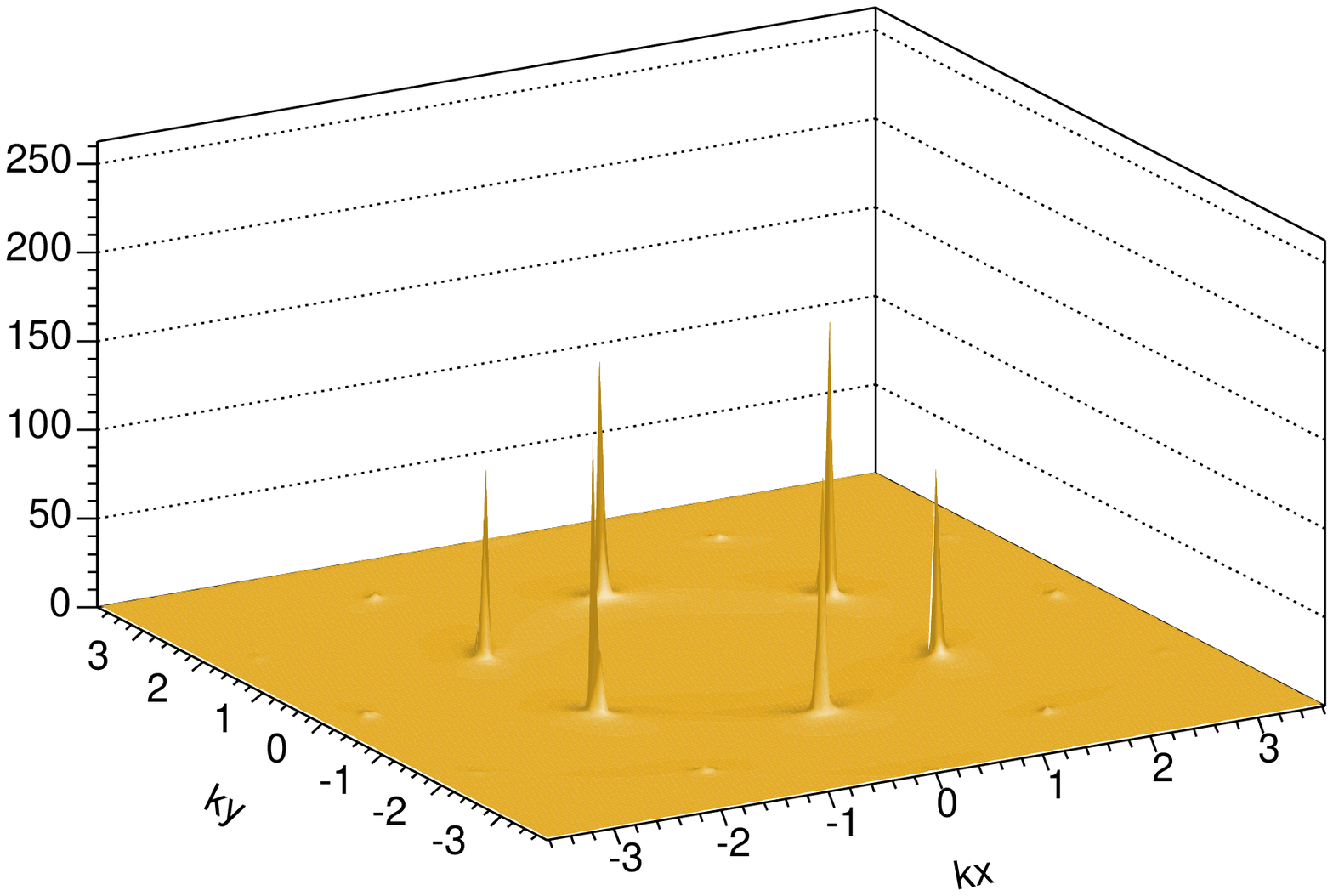} \\
   {\rm (c)}\qquad n_v=15                   & {\rm (d)}\qquad n_v=25                             
  \end{array}$
 \end{center}
 \caption{\label{sk} (color online). Static structure factor $S(\vec k)$ in the $k_x,k_y$ plane
          computed in a $^4$He 2D crystal at $\rho = 0.0765$\AA$^{-2}$ in a box with $M=1440$
          lattice sites and with different $n_v$.
          The wave vectors are in \AA$^{-1}$.}
\end{figure}

The double regime behavior of $\Delta E_{n_v}$ is reflected also in other properties of the 
system, for instance in the static structure factor $S(\vec k)$.
For a finite crystal the height $S_{\rm max}$ of the Bragg peaks has a strong dependence on
the number of particle $N$ and $S_{\rm max}/N$ shows a slow convergence (like $N^{-2/3}$ in 
3D\cite{Cepe3} and $N^{-1/2}$ in 2D) to the thermodynamic limit.
Note that being at $T=0$ K, the crystalline order is stable also in 2D.
Our computation of $S_{\rm max}/N$ in the perfect crystal verifies this $N^{-1/2}$ dependence as
expected for a 2D crystal.
This $N^{-1/2}$ dependence of $S_{\rm max}/N$ arises from missing phonon modes for $k<k_{\rm min}$
in a finite box so that essentially the same $N^{-1/2}$ contribution is expected to be present
in the defected system.
Therefore by taking the difference between $S_{\rm max}/N$ in the defected system and the value
in the prefect crystal such $N^{-1/2}$ contribution cancels out to a large extent so we can compare
results for crystal of different sizes and $n_v$.
Since the crystalline lattice relaxes around a vacancy, one expects that the heights of the Bragg
peaks in $S(\vec k)$ decrease with increasing number of vacancies.
This is observed, as shown in Fig.~\ref{f:str}, only when $n_v$ is small.
For large $n_v$ values we find a very different behavior.
The main Bragg peak height has some broadening but its integrated intensity\cite{nota3} does not 
show a significant dependence on $n_v$, rather it oscillates around an almost constant value.
In Fig.~\ref{sk} we show the static structure factor $S(\vec k)$ in the $\vec k$ plane obtained 
for $n_v=0$, 5, 15 and 25 in the largest considered box with $M=1440$ lattice sites.
When the number of vacancies is small (Fig.~\ref{sk}a) we find Bragg peaks in the first 
reciprocal vector star as sharp as in the case of the perfect crystal, within the resolution 
$\Delta k = \frac{2\pi}{L}$ due to the finite size of the simulation box, but with a decreased 
height.
Otherwise, see Fig.~\ref{sk}c and d, for large $n_v$, the Bragg peaks turn out to be slightly 
broadened, but their integrated intensity (see Fig.~\ref{f:str}) shows no significant dependence 
on $n_v$.
The broadening $\sigma_{\rm max}$ of the main Bragg peak has a linear dependence on the
concentration of defects $x_v$, with $\sigma_{\rm max} \simeq 1.5 \times x_v $ \AA$^{-1}$.

\begin{figure}
 \begin{center}
  \includegraphics[width=8.5cm]{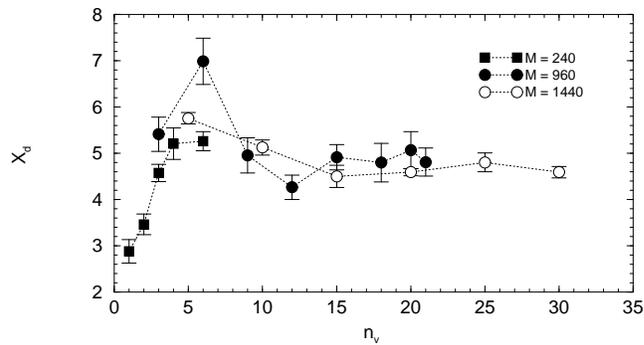}
 \end{center}
 \caption{\label{f:str2} Amount of disorder $X_d$ as a function of $n_v$.
          Lines are aid to the eyes.}
\end{figure}
\begin{figure}
 \begin{center}
  \includegraphics[width=12.0cm]{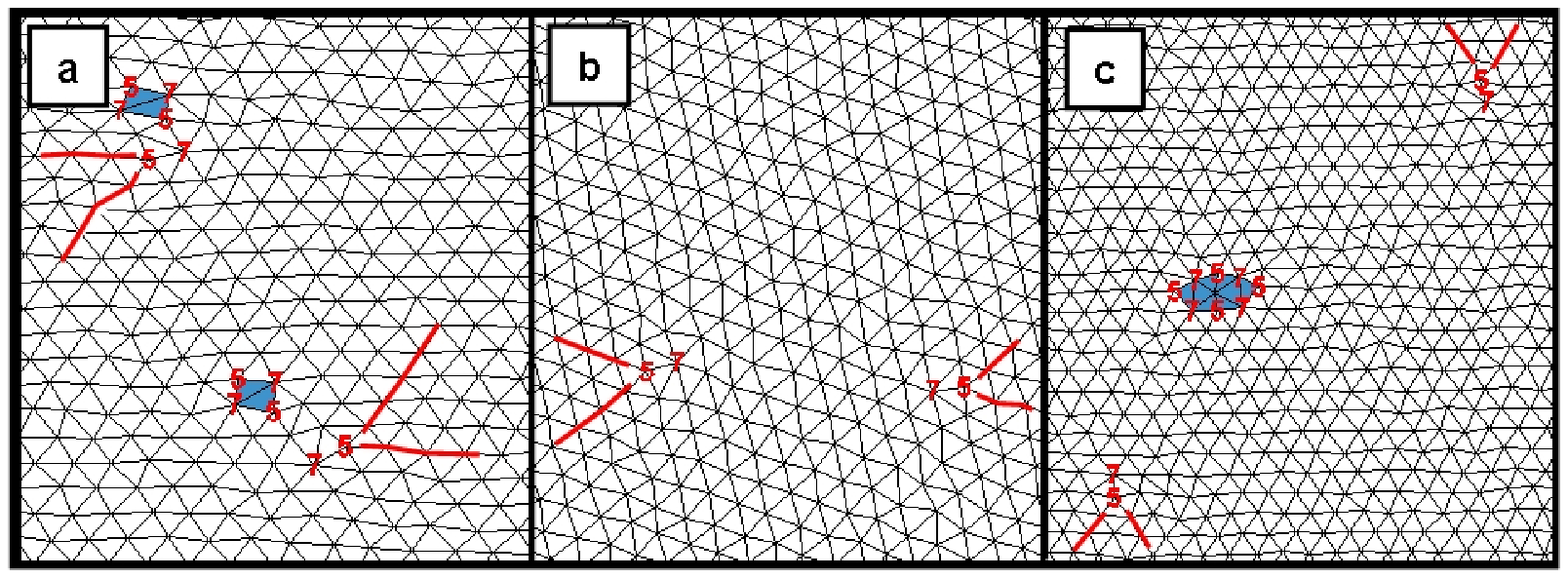}
 \end{center}
 \caption{\label{f:vac} (color online). Delaunay triangulation of typical equilibrium 
          configurations of a 2D $^4$He crystal at $\rho = 0.0765$\AA$^{-2}$ with 
          (a) $n_v=10$ and $M=480$, (b) $n_v=10$ and $M=490$ and (c) $n_v=20$ and $M=1440$. 
          The bold(red) lines indicate the interrupted lines of atoms.
          We report the coordination number for atoms only when different from 6.
          The couple 5--7 indicates a dislocation core in 2D and the clumps of two 5 and two 7
          (shadowed area in panel a) indicate bound pairs of dislocations.\cite{Ches2}
          In (c), besides the dislocations, also two close threefold symmetric 
          monovacancies\cite{Pres} are present (shadowed area).
          Notice that in all panels only a small part of the full system is shown.} 
\end{figure}
\begin{figure}
 \begin{center}
  \includegraphics[width=12.0cm]{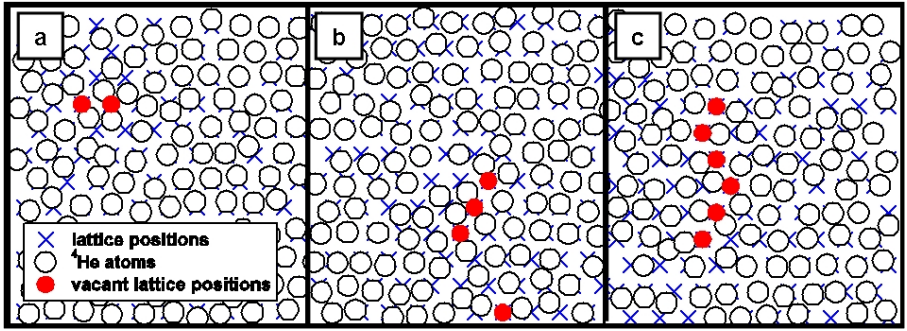}
 \end{center}
 \caption{\label{f:vac2} (color online). Vacancy configuration in a 2D $^4$He with $M=240$ and
          (a) $n_v=2$, (b) $n_v=4$ and (c) $n_v=6$.
          Vacancy positions are obtained as in Ref.~\cite{Ross2}. %
          Notice that in all panels only a small part of the full system is shown.}
\end{figure}
\begin{figure}
 \begin{center}
  \includegraphics[width=12.0cm]{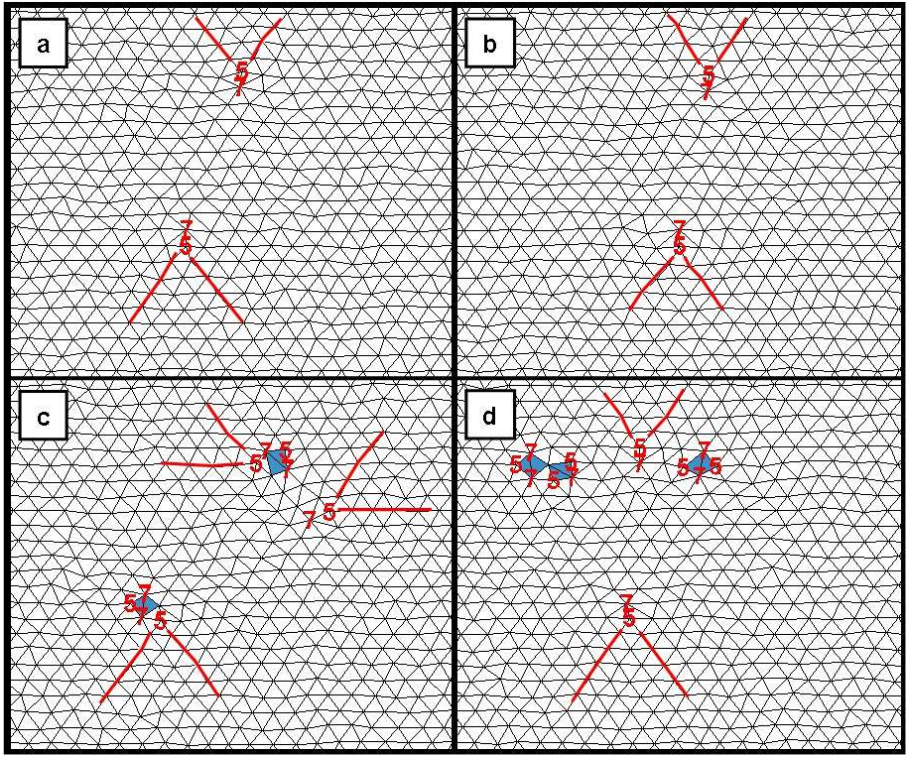}
 \end{center}
 \caption{\label{dislo} (color online). Example of the evolution of DT at different Monte Carlo 
          times (MCS) in a converged computation for a 2D $^4$He crystal at 
          $\rho = 0.0765$\AA$^{-2}$ with $n_v=10$ and $M=480$.
          The bold (red) lines indicate the interrupted lines of atoms.
          We report the coordination number for atoms only when different from 6.
          The couple 5--7 indicates a dislocation core in 2D.
          The clump of two 5 and two 7 (shadowed areas) indicates a bound pair of dislocations
          (BPD).\cite{Ches2}
          Starting from the configuration shown in the panel a, after 300 MCS dislocation cores 
          are found to be glided along the horizontal lattice axis by a lattice parameter (b).
          After 1100 MCS (c) we find three dislocation cores.
          Then after 1700 MCS we find again two dislocation cores, aligned along the vertical 
          simulation box axis, plus three BPD.
          In all panels the whole system is shown.}
\end{figure}

A similar behavior is displayed also by the amount of disorder $X_d$, plotted in Fig.~\ref{f:str2}.
$X_d$ increases with $n_v$ for small $n_v$ values and saturates to an almost constant value for
larger $n_v$.
By examining sampled configurations of the system in this regime, we find that it is no more 
possible to identify vacancy positions.
Rather one finds dislocations, as shown by a DT of the position of the atoms, see 
Fig.~\ref{f:vac}a and b.
Dislocation cores in 2D are point defects characterized by a couple 5--7 in the coordinations 
number of neighboring atoms in the DT.
We find that dislocations do not prefer any of the principal lattice directions.
Studying a large collection of independent configurations one finds that a large fraction of them
contains a couple of dislocation cores (see Fig.~\ref{f:vac}).
For example, for $n_v=20$ in the $M=960$ box, there is a probability of about 90\% for two 
dislocation cores and 10\% of finding more than two dislocation cores.
In addition there is a probability around 3\% of finding also some isolated vacancies besides 
dislocations (see Fig.~\ref{f:vac}c).
Similar values are found for $n_v=20$ in the $M=1440$ box.
We stress that, even if initially placed in a compact cluster configuration, vacancies do not 
separate into a vacancy rich region surrounded by a regular crystal. 
Rather they mostly reorganize themselves across the system giving rise to dislocations.
In this regime of large $n_v$ the quantity that maintains direct physical meaning is $N=M-n_v$, 
the number of atoms.
When atoms rearrange themselves giving rise to dislocations, the number of lattice sites becomes
{\it ill defined} because the lattice positions of the perfect crystal compatible with the 
simulation box and the pbc do no more represent equilibrium positions for the particles, and it is 
no more possible to recognize a regular lattice describing the equilibrium positions of the particles.
Still we continue to use $M$ and $n_v=M-N$ as a convenient measure of deviation 
from the ideal crystal $N=M$ where crystalline order is perfect.
In the regime of small $n_v$, vacancies are strongly correlated and prefer to form linear 
configurations, as shown in Fig.~\ref{f:vac2}.
These linear configurations can be considered as forerunners of the quantum dislocations found at
larger $n_v$.
This behavior has some similarities with the behavior of colloidal crystals in 
2D.\cite{Pres,Lech}
We find that dislocations are not fixed structures, rather they are very mobile and their number
changes along the simulation.
As an example, in Fig.~\ref{dislo} we show the evolution of the DT of a configuration
for $n_v=10$ vacancies in the box with $M=480$ sampled at different Monte Carlo times.

\begin{figure}
 \begin{center}
  \includegraphics[width=8.5cm]{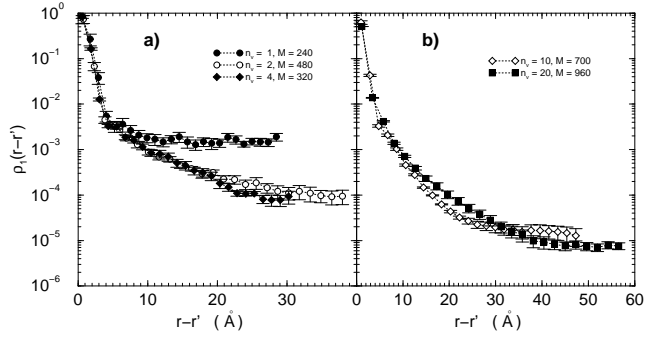}
 \end{center}
 \caption{\label{f:r1pv} One--body density matrix $\rho_1$ computed along the first neighbor 
          direction in 2D $^4$He crystals at $\rho = 0.0765$\AA$^{-2}$ in different boxes in 
          (a) vacancy regime and (b) dislocation regime.}
\end{figure}
\begin{figure}
 \begin{center}
  \includegraphics[width=12.0cm]{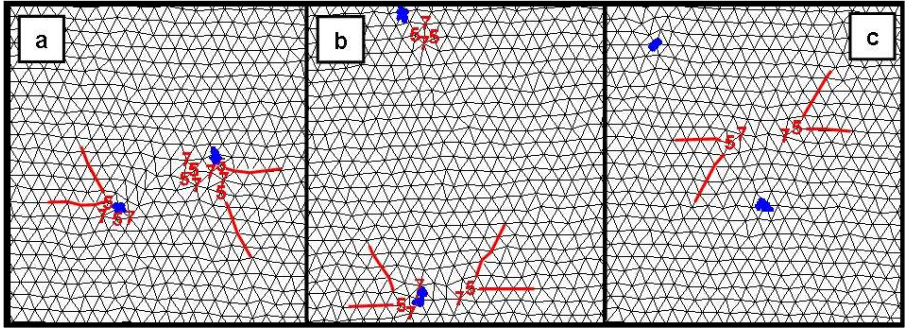}
 \end{center}
 \caption{\label{f:vac3} (color online). Delaunay triangulation of typical equilibrium
          configurations of a 2D $^4$He crystal at $\rho = 0.0765$\AA$^{-2}$ in an off--diagonal
          SPIGS simulation with $n_v=20$ and $M=960$.
          The bold (red) lines indicate the interrupted lines of atoms.
          We report the coordination number for atoms only when different from 6.
          The couple 5--7 indicates a dislocation core in 2D.
          The clump of two 5 and two 7 indicates a bound pair of dislocations.\cite{Ches2}
          The blue dots are the snapshots of the half--polymers.\cite{spigs}
          The two half--polymers are mainly found to occupy the dislocation cores like in (a),
          but they have been found also outside the core (both half--polymers in (c) or only 
          one in (b)).
          Notice that in all panels only a small part of the full system is shown.}
\end{figure}
\begin{figure}
 \begin{center}$
  \begin{array}{cc}
   \includegraphics[width=8.5cm]{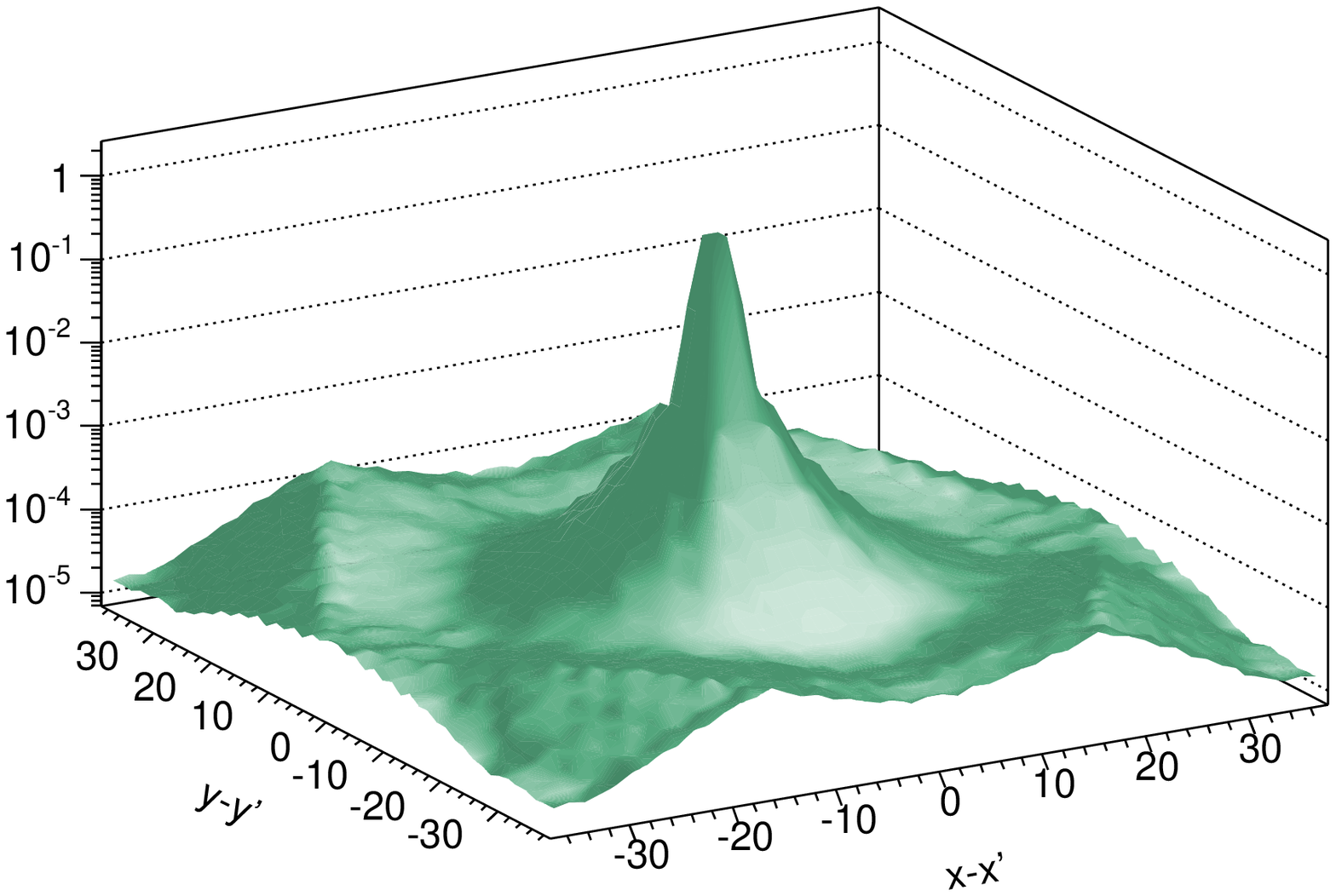} & \includegraphics[width=8.5cm]{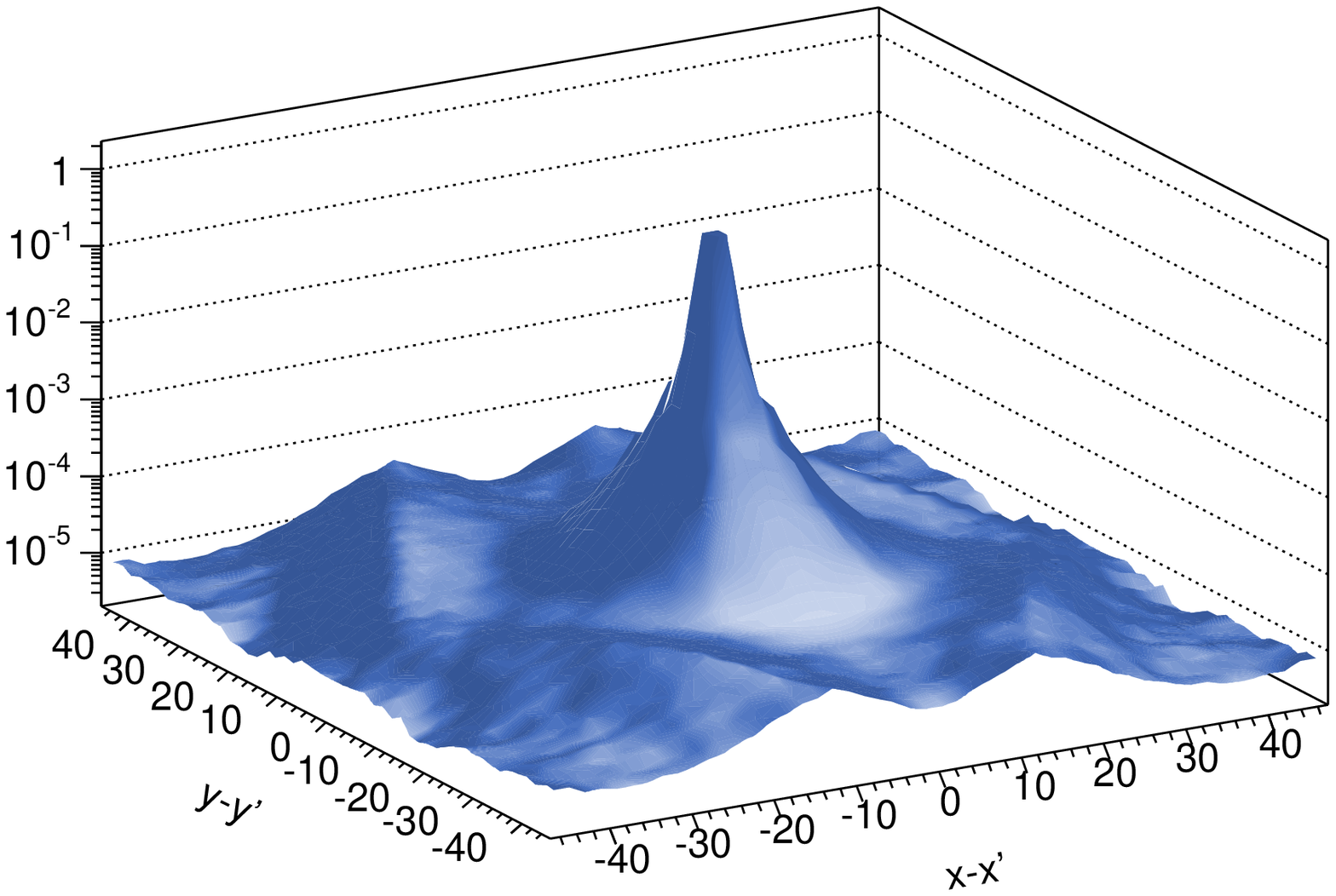} \\
   {\rm (a)}\qquad n_v=10, M=480            & {\rm (b)}\qquad n_v=10, M=700             \\
   \includegraphics[width=8.5cm]{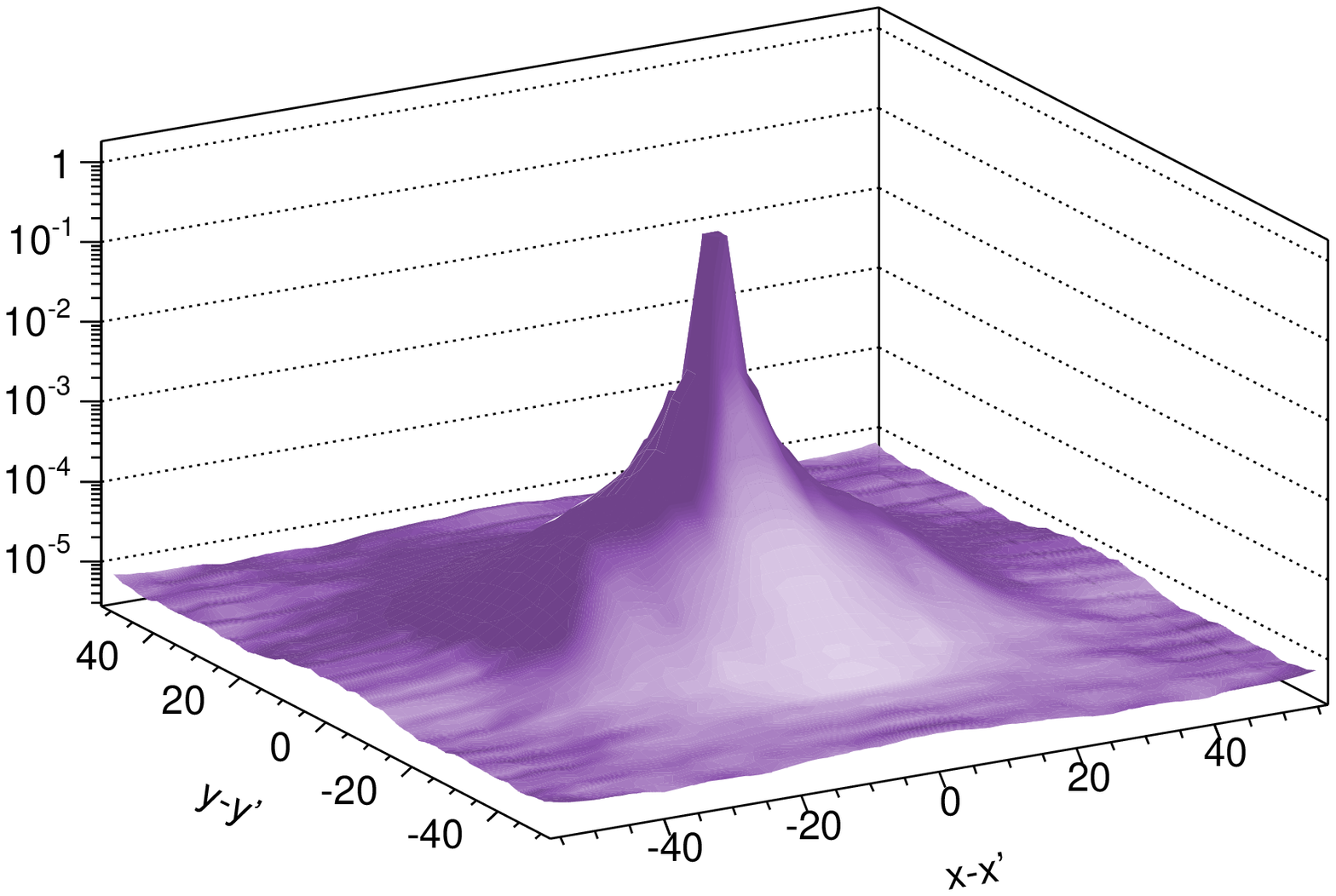}& \\
   {\rm (c)}\qquad n_v=20, M=960             & \\
  \end{array}$
 \end{center}
 \caption{\label{odlro} (color online). One--body density matrix $\rho_1(\vec{r},\vec{r}\,')$
          in the $x-x',y-y'$ plane computed in a defected $^4$He 2D crystal at
          $\rho = 0.0765$\AA$^{-2}$.}
\end{figure}

The one--body density matrix $\rho_1(\vec{r},\vec{r}\,')$ has a central role since the 
presence of a plateau at large $|\vec{r}-\vec{r}\,'|$ indicates the presence of ODLRO, i.e.
the system has Bose--Einstein condensation (BEC).
Even $\rho_1(\vec{r},\vec{r}\,')$ seems to show the signature of a double regime behavior
depending on the number of vacancies.
Some of our results for the one--body density matrix are reported in Fig.~\ref{f:r1pv}.
As found in the 3D case\cite{Gall}, when only a single vacancy is present $\rho_1$ displays a
non--zero plateau, which is the signature of ODLRO.
As vacancies are added, the tail of one--body density matrix is depressed, even if it
does not show a simple exponential decay. 
In order to be conclusive on the behavior of $\rho_1$ in the large distance range one should 
compute it in larger systems; we have faced these computationally intensive calculations in 
the $n_v$ range where vacancies give rise to quantum dislocations. Here (see Fig.~\ref{f:r1pv}b 
and Fig.~\ref{odlro}c) $\rho_1$ displays a recognizable plateau at large distances (very evident
in the largest system).
In order to understand the origin of the observed off--diagonal contributions to $\rho_1$ in the 
dislocation regime it is useful to recall that in a SPIGS computation $\rho_1(\vec{r},\vec{r}\,')$ 
(which is the probability amplitude to destroy a particle in $\vec{r}\,'$ and to create one in 
$\vec{r}$) is obtained by splitting one of the linear polymers in two half polymers,\cite{spigs} 
one departing from $\vec{r}$ and the other from $\vec{r}\,'$.
We find that the main contribution to the plateau comes from configurations where at least one of
the half--polymers occupies a dislocation core, see Fig.~\ref{f:vac3}(a,b).
This means that the system is able to transfer particles from a quantum dislocation to another.
This is quite distinct from the observed superfluidity along the core of a screw
dislocation\cite{Boni2} in 3D, in fact in 2D a dislocation core is a point defect and not a
linear one as in 3D.
Furthermore, the possibility of finding a half--polymer out of the cores like in
Fig.~\ref{f:vac3}(b,c) means that quantum dislocations are able also to induce vacancies in the
surrounding crystal.
Contrarily to diagonal properties, in the computed one--body density matrix we find
presence of significant size effects, as can be inferred for example by
the significant difference between the $\rho_1$ in the  $M=700$, $n_v=10$ system and in the
$M=960$, $n_v=20$ one, in the intermediate distance range ($20-40$\AA).
This is shown more clearly, in Fig.~\ref{odlro} where we report $\rho_1$ in the full plane,
one can notice the presence of ridges in the tail of $\rho_1$ in the smaller systems ($M=480$ 
and $M=700$, see panels (a) and (b)) that are no more present in the largest system (see 
Fig.~\ref{odlro}c).
Such ridges are due to a commensuration effect with the simulation box.
We have evidence that in the $M=480$ and $M=700$ systems dislocation cores
prefer to stay at a distance of about
$(n_v\pm 2)\times a$ ($a$ is the lattice parameter) that turns out to be comparable with $L_x/2$.
This is no more true in the box with $M=960$, in which case $\rho_1$ has no ridge in the tail
(see Fig.~\ref{odlro}c).

In order to determine how the plateau of $\rho_1$ is affected by finite size effects one 
should perform computations for an even larger number of particles, bur $N\simeq10^3$ is the
maximum number that presently we can handle.
Another reason for studying larger systems is to establish how the number of dislocations 
scales with the size of the system and with the number of vacancies.
Presumably supersolidity and BEC are present in a macroscopic system only if the concentration
of dislocations is finite.
In addition there is the question if the projection time $\tau$ is large enough to
have convergence to the exact result.
On the basis of indirect tests performed on diagonal observables which depend on 
long--range correlations (see Fig.~\ref{f:lr}) we are rather confident that the used value
of $\tau$ is large enough to get convergence, but we are not able to provide a direct test
by using a larger value of $\tau$ again for computational limitations (a computation with
$\tau=0.775$K$^{-1}$ with $\delta\tau=1/40$K$^{-1}$ and $10^3$ atoms has to handle about
$2\times10^5$ degrees of freedom).
Our results in the dislocation regime hint that ODLRO is present, but further tests are needed 
in order to firmly establishing this.
In any case our results on off--diagonal properties in 2D systems enlighten a novel ability 
of dislocations in inducing exchanges even in their surroundings.

\section{CONCLUSIONS}
\label{sec:conc}

In summary, we find that crystalline order in 2D $^4$He is stable also in presence of a large 
number of vacancies and there is no tendency to phase separation. 
This is true also in 3D, we have studied up to 2548 particles and 98 vacancies and in all cases 
we find that crystalline order is present, as shown by the presence of Bragg peaks in the static
structure factor.
When, in the 2D system, the number of vacancies is of order of 10 and above, vacancies inserted 
in the initial configuration loose their identity and most of them become quantum dislocations.
Such dislocations turn out to be very mobile and are able to induce exchange of particles across 
the system, which is necessary to supersolidity.
The ability of dislocations to induce vacancies in the surrounding crystal could be relevant also
for the 3D case.
If this feature will be confirmed in 3D, the superfluidity would be not restricted to the 
dislocation cores,\cite{Boni2} but exchange processes necessary for supersolidity would be
promoted by vacancies even in the bulk crystal far from dislocation cores.
Simulations with numbers of particle orders of magnitude larger are needed to answer the question 
whether the number of dislocations cores will increase with the system size and with the number 
of vacancies to give a finite concentration of dislocations able to induce ODLRO even in a 
macroscopic 2D $^4$He crystal, but this is actually beyond our computational possibilities and we
have to leave it for future investigations.
Moreover, how much the phase correlation triggered by dislocations depends on the concentration
of dislocations and its relevance for the 3D solid $^4$He are still open questions.
Our results mainly indicate that the usual concepts of commensurate or incommensurate solid, 
borrowed from  lattice models, is not appropriate for solid Helium, a system where the crystal 
lattice is not externally imposed, but is self-induced by the correlation among particles.
In fact, the number of lattice sites $M$ is found to be an ill defined quantity when the crystal
houses many vacancies/dislocations.

We do not address the origin (intrinsic or extrinsic) of the vacancies/dislocations studied in 
the present paper. 
The issue if the ground state could contain zero point defects is still 
debated.\cite{Gall,Rev,And2,Ross}
In case the ground state of solid $^4$He contains defects, our findings suggest that the 
Andreev--Lifshitz--Chester scenario\cite{Andr,Ches} would in a certain sense revive, but in terms
of {\it ground state dislocations} rather than vacancies.
On the other hand, even if the ground state has no zero point defects, our findings suggest that
in presence of extrinsic disorder this will manifest more in terms of fluctuating dislocations 
rather than vacancies, at least at low $T$ and in 2D.

This work was supported by the INFM Parallel Computing Initiative and by the Supercomputing 
facilities of CILEA.
M. Rossi would thank W. Lechner for useful discussions.


\begin{thebibliography}{99}
\bibitem{nota} Both $e(M-n_v)$ and $e(M)$ are ground state energies of two periodically repeated 
               small systems, and their difference $\Delta E_{n_v}$ has been used to estimate the
               formation energy of extra vacancies in bulk, therefore $\Delta E_{n_v}$ is a derived 
               quantity.\cite{Gall}
\bibitem{Dela} The Delaunay triangulation (DT) for a set $P$ of $N$ points in the plane is a 
               triangulation $\mathcal T(P)$ such that no point in $P$ is inside the 
               circumcircle of any triangle in $\mathcal T(P)$.
               DT corresponds to the dual graph of the Voronoi tessellation, i.e. the 
               partitioning of the plane into convex polygons such that each polygon contains 
               exactly one point in $P$ and every point of the plane closer to it than to any 
               other point of $P$.
               Then each point in $P$ is linked only to its nearest neighbors.
               The number of the sides of Voronoi polygons (of the links in DT) is the 
               coordination number of the generating point.
\bibitem{nota2} The algorithm proposed in Ref.~\cite{Cepe2} by construction provides always %
                $n_v=M-N$ vacancy positions.
                For few vacancies the results with this algorithm agree with those obtained with the
                coarse graining method used here.
                When many vacancies are present (and turns into dislocations) the vacancy positions 
                given by the algorithm of Ref.~\cite{Cepe2} are extremely erratic from one %
                configuration to the following one and thus are not reliable.
\bibitem{nota3} The broadening of the Bragg peaks occurs on a small number of close points in the 
                $k$-grid (wave vectors in a finite system with pbc are quantized), thus we refer to 
                integrated intensity as the sum of $S(\vec k)$ over the $k$-points contributing to 
                the Bragg peak.
\end{thebibliography}
\end{document}